\documentclass[12pt]{article}

\usepackage{amssymb}
\usepackage{amsfonts}

\begin{document}

\title{$q$-Deformation and free statistics \\ for interaction of a field and a particle}

\author{S.V. Kozyrev\footnote{Steklov Mathematical Institute of Russian Academy of Sciences, {\tt kozyrev@mi-ras.ru}}}

\maketitle

\begin{abstract}
Emerging of free (or quantum Boltzmann) statistics for a model of quantum particle interacting with quantum field is described in the stochastic limit without dipole approximation. The quantum field is considered in a Gaussian (for example temperature) state. Entangled operators which describe interaction of the field and the particle satisfy the $q$-deformed relations which in the stochastic limit generate free statistics.
\end{abstract}

\section{Introduction}

In the present paper we discuss the stochastic limit for a quantum particle interacting with a quantum Bose field without dipole approximation. We shaw that in this limit some collective excitations of the field and the particle (the entangled operators defined as a free evolution of contributions to the interaction Hamiltonian of the field and the particle) obtain free (or quantum Boltzmann) statistics. The algebra of free creation--annihilation operators is given by the relations
$$
b_ib_j^{\dag}=\delta_{ij},
$$
annihilation operators $b_i$ do not commute and constitute a free algebra. We show that in the model under consideration some deformation of this algebra emerges. Free operators of creation and annihilation were studied in the free probability theory, see \cite{Voiculescu}, these operators are related to the theory of random matrices and the limit semicircle Wigner law.

The stochastic limit studies the dynamics of quantum system interacting with environment with coupling constant $\lambda$ in the van Hove--Bogoliubov rescaling $t\mapsto t/\lambda^2$ in the limit $\lambda\to0$, i.e. the effect of weak interactions in the regime of large times  \cite{theBook}.

Before the stochastic limit the entangled operators constitute an algebra with ''dynamically'' $q$-deformed relations, see formulae (\ref{qdef1}), (\ref{qdef2}), (\ref{qdef3}) below, free creation and annihilation operators arise as the stochastic limit of the $q$-deformed creations and annihilations. The quantum field (the environment) is in some Gaussian (for example temperature) state.
For the case when the environment field is in the Fock state (i.e. the state is given by the vacuum average) the corresponding model was considered in \cite{qdef}, see also \cite{theBook}, \cite{hotfree}. Other models with non-linear interaction and free statistics in the stochastic limit were studied in \cite{120}, \cite{129}. In the present paper we generalize the approach of \cite{qdef} for the case of general Gaussian state. For the consideration of the temperature state we apply the free analogue of the temperature double construction, i.e. a variant of the Bogoliubov  transformation which allows to represent an arbitrary Gaussian state of the field using Fock states for a pair of auxiliary fields (see section 4 of this paper).

The exposition of the present paper is as follows. In Section 2 the model of interaction of a particle and a Bose field without dipole approximation is described, entangled operators describing collective excitations of the field and the particle are introduced, dynamically $q$-deformed relations for these operators are described, and the example of 4-point correlation function demonstrates the emerging of free statistics in the stochastic limit. In Section 3 correlation functions for the algebra of $q$-deformed relations for entangled operators are computed in an arbitrary (non-squeezed mean zero) Gaussian state and the stochastic limit for these correlation functions is obtained. In Section 4 it is shown that the stochastic limit of correlation functions of Section 3 equals to correlation functions for some algebra of free creation and annihilation operators constructed in analogy to the construction of the temperature double.

\section{Dynamical $q$-deformation}

\noindent{\bf Formulation of the model.}\quad
Let us consider Hamiltonian of a quantum particle interacting with quantum Bose field without dipole approximation
$$
H=H_0+\lambda H_I=\int\omega(k)a^{\dag}(k)a(k)dk+{\frac{1}{2}}\,p^2+\lambda H_I,
$$
$$
H_I=p\cdot{\cal A}(q)+{\cal A}(q)\cdot p=\int d^3k\left(g(k) p\cdot e^{ikq} a^{\dag}(k)+ \overline {g(k)} p\cdot e^{-ikq}  a(k)\right) + h.c.,
$$
$$
q=(q_{1}, q_{2}, q_{3}),\quad p=(p_{1}, p_{2}, p_{3}),\quad [q_{h}, p_{k} ] = i \delta_{hk},
$$
$$
a(k)=(a_{1}(k), a_{2}(k), a_{3}(k)),\quad a^{\dag}(k)=
(a^{\dag}_{1}(k), \ldots ,a^{\dag}_3(k)),\quad [a_{j}(k),a_{h}^{\dag}(k')]=\delta_{jh}\delta(k-k').
$$

Here $q$ and $p$ are coordinate and momentum of the particle, the field $a(k)$ possesses dispersion $\omega(k)$, function $g(k)$ is a formfactor of interaction of the field and the particle.

In the procedure of the stochastic limit one computes approximation for the perturbation series for evolution operator in interaction picture. We study the free evolution of the interaction Hamiltonian $H_I(t)=e^{it H_0}H_Ie^{-itH_0}$ and the van-Hove--Bogoliubov rescaling by square of the coupling constant $\lambda$, then we take the weak coupling limit $\lambda\to 0$
$$
{\frac{1}{\lambda}}\,H_I\left({\frac{t}{\lambda^2}}\right)=
{\frac{1}{\lambda}}A(t/\lambda^2) + h.c. =
\int d^3k p(\overline g(k) a_\lambda(t,k)+  g(k)a^{\dag}_\lambda(t,k)) + h.c.
$$

We obtain values which describe the simultaneous free evolution of the field and the system (entangled operators)
\begin{equation}\label{entangled1}
a_\lambda(t,k)={\frac{1}{\lambda}}\,e^{i{\frac{t}{\lambda^2}}\,H_0}e^{-ikq}a(k)e^{-i{\frac{t}{\lambda^2}}\,H_0}
=\frac{1}{\lambda}e^{-i{\frac{t}{\lambda^2}}\left[\omega(k)+{\frac{1}{2}}\,k^2+kp\right]}e^{-ikq}a(k),
\end{equation}
\begin{equation}\label{entangled2}
a^{\dag}_\lambda(t,k)=\frac{1}{\lambda}e^{i{\frac{t}{\lambda^2}}\left[\omega(k)-{\frac{1}{2}}\,k^2+kp\right]}e^{ikq}a^{\dag}(k).
\end{equation}

\medskip

\noindent{\bf Dynamical $q$-deformation.}\quad
There is an important observation: entangled operators satisfy the algebra of dynamically $q$-deformed relations
\begin{equation}\label{qdef1}
a_\lambda(t,k)a^{\dag}_\lambda(t',k')=a^{\dag}_\lambda(t',k')a_\lambda(t,k)
q_\lambda(t-t',kk')+
{\frac{1}{\lambda^2}}\,q_\lambda\left(t-t',\omega(k)+{\frac{1}{2}}\,k^2+kp\right)\delta(k-k'),
\end{equation}
\begin{equation}\label{qdef2}
a_\lambda(t,k)p=(p+k)a_\lambda(t,k),
\end{equation}
\begin{equation}\label{qdef3}
a_\lambda(t,k)a_\lambda(t',k')=a_\lambda(t',k')a_\lambda(t,k)q_\lambda^{-1}(t-t',kk'),
\end{equation}
where the function $q$ is the oscillating exponent
$$
q_\lambda(t-t',x)=e^{-i{\frac{t-t'}{\lambda^2}}\,x}.
$$

Let us call $t-t'$ the time argument of the oscillating exponent and $x$ the energetic argument. Let us consider the problem of construction of the master--field, i.e. the stochastic limit of the entangled operator
$$
b(t,k)=\lim_{\lambda\to0}a_\lambda(t,k)
$$
and the corresponding commutation relations.

We would like to use the limit for the oscillation exponent (in the space of tempered distributions)
\begin{equation}\label{oscillation}
\lim_{\lambda\to0}q_\lambda(t,x)=0,
\qquad\lim_{\lambda\to0}{\frac{1}{\lambda^2}}\,
q_\lambda(t,x)=2\pi\delta(t)\delta(x).
\end{equation}

It is natural to expect that commutation relations for the master--field are limits of the dynamically $q$-deformed relations (\ref{qdef1}), (\ref{qdef2})
\begin{equation}\label{free1}
b(t,k)b^{\dag}(t',k')=2\pi\delta(t-t')\delta\left(\omega(k)+{\frac{1}{2}}\,k^2+kp\right)\delta(k-k'),
\end{equation}
\begin{equation}\label{free2}
b(t,k)p=(p+k)b(t,k),
\end{equation}
and the third relation (\ref{qdef3}) generates the free statistics, i.e. makes creations non-commuting (contrary to the Bose case) and creations form a free algebra. We will check that the master--field has free statistics and satisfies (\ref{free1}), (\ref{free2}) in the case when the field $a(k)$ is in the Fock state. For more general Gaussian states of the field (in particular temperature state) we will get a generalization of (\ref{free1}), (\ref{free2}) --- a quantum algebra with free statistics and relations which depend on the temperature. Here master--field is delta--correlated with respect to time, i.e. it is a quantum white noise (with free statistics).

\medskip

\noindent{\bf Example: Free statistics for 4-point correlator.}\quad
Let us discuss the emerging of free statistics in the limit $\lambda\to0$. Operators $a_\lambda(t,k)$ satisfy relations (\ref{qdef3}). One could suggest that in the limit $\lambda\to0$ these relations tend to $b(t_1,k_1)b(t_2,k_2)=0$. But this is not the case. Actually this relation leads to free statistics (i.e. annihilation operators in the limit do not commute). To prove this let us consider the 4-point correlator (in the Fock state)
\begin{equation}\label{4-point}
\langle
a_\lambda(t_1,k_1)a_\lambda(t_2,k_2)
 a^{\dag}_\lambda(t'_2,k'_2)
a^{\dag}_\lambda(t'_1,k'_1)\rangle=
\end{equation}
$$
=\frac{1}{\lambda^2}\,q_{\lambda}\left(t_2-t'_2,
\omega(k_2)+\frac{1}{2}k_2^2+ k_2(p+k_1)\right)\delta(k_2-k'_2) $$ $$
\frac{1}{\lambda^2}\,q_{\lambda}\left(t_1-t'_1,
\omega(k_1)+\frac{1}{2}k_1^2+ k_1p\right)\delta(k_1-k'_1) q_{\lambda}(t_2-t'_2,k_2 k'_2)
+
$$
$$
+
\frac{1}{\lambda^2}\,q_{\lambda}\left(t_1-t'_2,
\omega(k_1)+\frac{1}{2}k_1^2+ k_1 p\right)\delta(k_1-k'_2) $$ $$
\frac{1}{\lambda^2}\,q_{\lambda}\left(t_2-t'_1,
\omega(k_2)+\frac{1}{2}k_2^2+ k_2 p\right)\delta(k_2-k'_1) q_{\lambda}(t_2-t'_2,k_2 k'_2).
$$

The first contribution contains the product of terms $\frac{1}{\lambda^2}q_{\lambda}(t_1-t'_1,\cdot)$, $\frac{1}{\lambda^2}\,q_{\lambda}(t_2-t'_2,\cdot)$ (which tend to $\delta$-functions), and $q_{\lambda}(t_2-t'_2,\cdot)$ (which tend to zero), but this term (tending to zero) has the time argument coinciding with the time argument of one of non-zero in the limit $\lambda\to 0$ oscillating exponents. Hence the product of these two terms survives in the limit (the limit of products is not equal to the product of limits).

In the second contribution all oscillating exponents have different time arguments, thus the tending to zero oscillating exponent can not cancel and the limit $\lambda\to 0$ of the second contribution equals to zero. The first contribution corresponds to a non-crossing, or rainbow diagram, the second contribution corresponds to a crossing diagram. One can see that in the stochastic limit only non-crossing diagrams survive.

Let us consider the 4-point correlator with permutated first two terms (using the commutation relation (\ref{qdef3}))
\begin{equation}\label{4-point'}
\langle
a_\lambda(t_2,k_2)a_\lambda(t_1,k_1)
a^{\dag}_\lambda(t'_2,k'_2)a^{\dag}_\lambda(t'_1,k'_1)\rangle
q_\lambda^{-1}(t_1-t_2,k_1k_2)=
\end{equation}
$$
=q_\lambda^{-1}(t_1-t_2,k_1k_2)\biggl(
\frac{1}{\lambda^2}\,q_{\lambda}\left(t_1-t'_2,
\omega(k_1)+\frac{1}{2}k_1^2+ k_1(p+k_1)\right)\delta(k_1-k'_2)
$$
$$
\frac{1}{\lambda^2}\,q_{\lambda}\left(t_2-t'_1,
\omega(k_2)+\frac{1}{2}k_2^2+ k_2 p\right)\delta(k_2-k'_1) q_{\lambda}(t_1-t'_2,k_1 k'_2)
+
$$
$$
+
\frac{1}{\lambda^2}\,q_{\lambda}\left(t_2-t'_2,
\omega(k_2)+\frac{1}{2}k_2^2+ k_2 p\right)\delta(k_2-k'_2)
$$
$$
\frac{1}{\lambda^2}\,q_{\lambda}\left(t_1-t'_1,
\omega(k_1)+\frac{1}{2}k_1^2+ k_1 p\right)\delta(k_1-k'_1) q_{\lambda}(t_1-t'_2,k_1 k'_2)
\biggr).
$$

In this case due to presence of the oscillating exponent $q_\lambda^{-1}(t_1-t_2,k_1k_2)$ this first contribution tends to zero. In the second contribution two oscillating exponents (which tend to zero) cancel and give $q_\lambda(t_2-t'_2,k_1k_2)$, the time argument is this term coincides with the argument in one of non tending to zero oscillating exponents. Hence application of relation (\ref{qdef3}) implies that in the limit of 4-point correlator crossing diagram survives and non-crossing vanishes. But these diagrams correspond to above diagrams in (\ref{4-point}) (for (\ref{4-point}) and (\ref{4-point'}) crossing and non-crossing diagrams change roles, hence if one tries to permute annihilations using relation (\ref{qdef3}), then operators in the limit permute restoring the initial order). Therefore commutation relation (\ref{qdef3}) in the limit vanishes (the statistics becomes free).

\section{Computation of $N$-point correlator}

In present Section we compute the form of many-point correlator for algebra of entangled operators (\ref{qdef1}), (\ref{qdef2}), (\ref{qdef3})
$$
\langle a^{\varepsilon_1}_\lambda(t_1,k_1)\dots a^{\varepsilon_N}_\lambda(t_N,k_N)\rangle.
$$
Here $a^{\varepsilon}$ denotes $a$ or $a^{\dag}$ $(\varepsilon=-1$ for $a$ and $\varepsilon=1$ for $a^{\dag}$), brackets $\langle \cdot \rangle$ denote a Gaussian (in particular temperature) state.

\medskip

\noindent{\bf Gaussian state} (non-squeezed and mean-zero) for the algebra of canonical commutation relations is described by the Wick theorem (or variation of this theorem by Bloch and de Dominicis). In this state average of a monomial over creations and annihilations can be non-zero only if the number $n$ of creations equals to the number of annihilations. Non-zero pairings of creations and annihilations (averages of products of two operators) are equal to
$$
\langle a^{\dag}(k)a(k')\rangle=N(k)\delta(k-k'),\qquad \langle a(k)a^{\dag}(k')\rangle=(N(k)+1)\delta(k-k').
$$

In particular for the temperature state
$$
N(k)=\frac{1}{e^{\beta\omega(k)}-1},
$$
where $\beta$ is the inverse temperature and $\omega(k)$ is the dispersion of the field with wave vector $k$ (the Planck constant is taken equal to one).

For a monomial of length $N=2n$ correlation function 
$$
\langle a^{\varepsilon_1}(k_1)\dots a^{\varepsilon_N}(k_N) \rangle,\quad N=2n
$$
is equal to the sum over partitions of the monomial in pairs of creations and annihilations of products of pairings and has the following form.

Let $\varepsilon=(\varepsilon_1,\dots,\varepsilon_{2n})$ be a sequence of creations and annihilations in the monomial, $\sigma(\varepsilon)=\{(m_j,m'_j)\}$, $j=1,\dots,n$ be a partition of the monomial in pairs of creations and annihilations ($m_{j}$ is the position of creation and $m'_{j}$ is the position of annihilation in the pair, all positions in the partition are different). Let us also introduce the symbol of ordering of a pair  $\delta_j={\rm sign}\left(m_{j}-m'_{j}\right)$ (i.e. $\delta_j=1$ if $a^\dag$ is on the right side of $a$ in the $j$-th pair and $\delta_j=-1$ in the opposite case). To any partition of a monomial one can put in correspondence a diagram where operators are vertices in a segment ordered in relation to their positions in the monomial (to the left in the monomial --- to the left in the segment), and pairs are shown by edges connecting vertices, let the segment belongs to the abscissa axis in the coordinate plane and the edges lie in the upper half-plane. A diagram is called non-crossing if edges can be depicted without crossings. Any vertex belongs to exactly one edge $(m_j,m'_j)$. Let us enumerate edges in the diagram according to order of their left vertices (i.e. edge with left vertex to the left will have smaller number in the numeration of edges).

The correlation function is equal to a sum over partitions $\sigma$ (i.e. over diagrams)
$$
\langle a^{\varepsilon_1}(k_1)\dots a^{\varepsilon_N}(k_N) \rangle=\sum_{\sigma(\varepsilon)}\prod_{j=1}^n M(k_{m_j})\delta(k_{m_j}-k_{m'_j}),
$$
\begin{equation}\label{M(k)}
M(k_{m_j})=N(k_{m_j})+\frac{1}{2}(\delta_j+1),\quad \delta_j={\rm sign}\left(m_{j}-m'_{j}\right).
\end{equation}

In particular for the Fock state $N(k)=0$ and $m_j>m'_{j}$ for all non-zero contributions to the correlation function. Due to presence of delta-functions wave vectors $k_{m_j}$, $k_{m'_j}$ of creations and annihilations in a pair coincide.

\medskip

\noindent{\bf $N$-point correlator for entangled operators}  (\ref{entangled1}), (\ref{entangled2}) has the form
$$
\langle a^{\varepsilon_1}_\lambda(t_1,k_1)\dots a^{\varepsilon_N}_\lambda(t_N,k_N)\rangle=
\frac{1}{\lambda^{2n}}\prod_{s=1}^{2n}\left[e^{-i{\frac{t_s}{\lambda^2}}\left[\omega(k_s)+{\frac{1}{2}}\,k_s^2+k_sp\right]}e^{-ik_sq}\right]^{-\varepsilon_s} \sum_{\sigma(\varepsilon)}\prod_{j=1}^n M(k_{m_j})\delta(k_{m_j}-k_{m'_j}),
$$
which contains the product of delta-functions corresponding to edges in the diagram (pairs of creations and annihilations from the partition of the monomial) and the product of exponents corresponding to vertices of the diagram. This product contains non-commuting operators and is ordered from the left to the right.

Let us also enumerate time arguments for entangled operators as $t_{m_j}$, $t_{m'_j}$. Analogously, let us enumerate indices $\varepsilon_{m_j}$, $\varepsilon_{m'_j}$. Both quadratic correlation functions (pairings of vertices of the diagram) $\langle a_\lambda(t_{m'_j},k_{m'_j})a^{\dag}_\lambda(t_{m_j},k_{m_j})\rangle$, $\langle a^{\dag}_\lambda(t_{m_j},k_{m_j})a_\lambda(t_{m'_j},k_{m'_j})\rangle$ corresponding to edges of the diagram can be put in the form
\begin{equation}\label{pairing}
\frac{1}{\lambda^{2}}e^{i{\frac{\left(t_{m_{j}}-t_{m'_{j}}\right)}{\lambda^2}}\left[\omega(k_{m_{j}})+\delta_j\frac{1}{2}k_{m_{j}}^2+pk_{m_{j}}\right]}
M(k_{m_{j}})\delta(k_{m_j}-k_{m'_j}).
\end{equation}

Let us express $N$-point correlator using above correlation functions.

Correlation function is a sum over diagrams (partitions of the monomial). We would like to express contribution to the correlation function which corresponds to a diagram in the form of product over edges of pairings of vertices, this product should also contain multipliers arising from permutations of the terms in the products of non-commuting exponents in the above formula. To put contribution in this form one has to move multipliers  $e^{i \delta_l k_l q}$ in the product (corresponding to vertices connected by edge) from the right to the left (move the right end of the edge to the left end). Since edge corresponds to delta-function of wave vectors, multipliers $e^{\pm i kq}$ for different ends of the edge cancel. We obtain product of commuting operators (which contain functions of the momentum $p$).

Let us introduce notations $a(l)=\min(m_{l},m'_{l})$, $b(l)=\max(m_{l},m'_{l})$. Let us perform the described above procedure of moving multipliers for the first pair $(m_{1},m'_{1})$. We will obtain pairing (\ref{pairing}) for the first pair and commutation of multipliers gives (for $l=1$)
$$
\prod_{s=a(l)+1}^{b(l)-1}\left[e^{-i{\frac{t_s}{\lambda^2}}\left[\delta_l k_s k_{m_l}\right]}\right]^{-\varepsilon_s} = \prod_{s=a(l)+1}^{b(l)-1}e^{i{\frac{t_s}{\lambda^2}}\varepsilon_s\delta_l k_s k_{m_l}}.
$$
Repeating this procedure for all edges of the diagram we get for contribution to the correlation function related to diagram the expression 
$$
\prod_{j=1}^{n}\frac{1}{\lambda^{2}}e^{i{\frac{\left(t_{m_{j}}-t_{m'_{j}}\right)}{\lambda^2}}\left[\omega(k_{m_{j}})+\delta_j\frac{1}{2}k_{m_{j}}^2+pk_{m_{j}}\right]}
M(k_{m_{j}})\delta(k_{m_j}-k_{m'_j})
\prod_{s=a(j)+1}^{b(j)-1}e^{i{\frac{t_s}{\lambda^2}}\varepsilon_s\delta_j k_s k_{m_j}}.
$$

Herewith if some $l$-th edge is situated inside the $j$-th edge then the product over $s$ will contain two contributions from the $l$-th edge (corresponding to ends of this edge), moreover wave vectors for ends of the edge will coincide. Hence product of these terms is equal to 
$$
e^{i{\frac{t_{a_l}-t_{b_l}}{\lambda^2}}\varepsilon_{a_l}\delta_j k_{m_l} k_{m_j}} = e^{i{\frac{t_{m_l}-t_{m'_l}}{\lambda^2}}\delta_j k_{m_l} k_{m_j}},
$$
therefore this contribution will depend on the time argument which coincides with the time argument for some edge of the diagram. If some edges cross we will obtain time arguments which will not correspond to any edge of the diagram. In the stochastic limit $\lambda\to0$ contributions of these diagrams will tend to zero. Product of contributions of crossings of edges in the diagram which cross the $j$-th edge takes the form 
$$
\left[\prod_{l:a(l)<a(j)<b(l)<b(j)} e^{i{\frac{t_{b_l}}{\lambda^2}}\varepsilon_{b_l}\delta_j k_{b_l} k_{m_j}}\right]
\left[\prod_{l:a(j)<a(l)<b(j)<b(l)} e^{i{\frac{t_{a_l}}{\lambda^2}}\varepsilon_{a_l}\delta_j k_{a_l} k_{m_j}}\right].
$$

Summing up the above discussion and using (\ref{oscillation}) we get:

\medskip

\noindent{\bf Theorem 1}\quad {\sl
$N$-point correlation function of entangled operators has the form
\begin{equation}\label{N-point}
\langle a^{\varepsilon_1}_\lambda(t_1,k_1)\dots a^{\varepsilon_N}_\lambda(t_N,k_N)\rangle= $$ $$=
\sum_{\sigma(\varepsilon)}\prod_{j=1}^n
\frac{1}{\lambda^{2}}e^{i{\frac{\left(t_{m_{j}}-t_{m'_{j}}\right)}{\lambda^2}}\left[\omega(k_{m_{j}})+\delta_j\frac{1}{2}k_{m_{j}}^2+pk_{m_{j}}
+\sum_{l:a(l)<a(j)<b(j)<b(l)} \delta_l k_{m_l}k_{m_{j}}\right]}
$$
$$
\left[\prod_{l:a(l)<a(j)<b(l)<b(j)} e^{i{\frac{t_{b_l}}{\lambda^2}}\varepsilon_{b_l}\delta_j k_{b_l} k_{m_j}}\right]
\left[\prod_{l:a(j)<a(l)<b(j)<b(l)} e^{i{\frac{t_{a_l}}{\lambda^2}}\varepsilon_{a_l}\delta_j k_{a_l} k_{m_j}}\right]
$$
$$
M(k_{m_j})\delta(k_{m_j}-k_{m'_j}).
\end{equation}

Here $a(l)=\min(m_{l},m'_{l})$, $b(l)=\max(m_{l},m'_{l})$, $M(k_{m_j})$ is given by (\ref{M(k)}).

The stochastic limit of this correlator takes the form
\begin{equation}\label{N'-point}
\lim_{\lambda\to 0}\langle a^{\varepsilon_1}_\lambda(t_1,k_1)\dots a^{\varepsilon_N}_\lambda(t_N,k_N)\rangle= $$ $$=
\sum_{\sigma'(\varepsilon)}\prod_{j=1}^n 2\pi
\delta \left(t_{m_{j}}-t_{m'_{j}}\right)\delta\left[\omega(k_{m_{j}})+\delta_j\frac{1}{2}k_{m_{j}}^2+pk_{m_{j}}
+\sum_{l:a(l)<a(j)<b(j)<b(l)} \delta_l k_{m_l}k_{m_{j}}\right]
$$ $$
M(k_{m_j}) \delta(k_{m_j}-k_{m'_j}),
\end{equation}
where notation $\sigma'(\varepsilon)$ means that the summation is restricted to non-crossing diagrams.

}

\section{Algebra of the master--field}

\noindent
{\bf Construction of the temperature double} (a variant of the Bogoliubov transformation) allows to express arbitrary Gaussian state using the Fock state. Let us consider for bosonic algebra transformation which expresses annihilation operator $a(k)$ in the form of linear combination of creation and annihilation $a_1(k)$ and $a_2^{\dag}(k)$ for two independent Bose fields (here $u$, $v$  are number parameters)
$$
a(k)\mapsto u a_{1}(k)+ v a_2^{\dag}(k),
$$
$$
a^{\dag}(k)\mapsto u^* a^{\dag}_{1}(k)+ v^* a_2(k).
$$

Commutation relation conserves if 
$$
[u a_{1}(k)+ v a_2^{\dag}(k),u^* a^{\dag}_{1}(k)+ v^* a_2(k)]=|u|^2-|v|^2=1.
$$

Let fields $a_1(k)$, $a_2(k)$ be in Fock states. Then
$$
\langle a^{\dag}(k)a(k')\rangle=|v|^2\delta(k-k'),
$$
$$
\langle a(k)a(k')\rangle=\langle a^{\dag}(k)a^{\dag}(k')\rangle=0.
$$

Hence using this transformation the Fock state becomes a Gassian non-Fock state (non-squeezed mean zero). To prove this statement it is sufficient to show that correlation functions for the field $a(k)$ can be computed using the Wick theorem which is guaranteed by the Wick theorem for fields $a_1(k)$, $a_2(k)$.

The next theorem allows to represent stochastic limits of correlation functions computed in Theorem 1 of previous Section using analogue of the temperature double construction for the algebra of master--field (quantum noise) with free statistics. Contrary to the bosonic case for the free temperature double not only the state changes but also commutation relations for the master--field.

\medskip

\noindent{\bf Theorem 2}\quad {\sl
The stochastic limit of the $N$-point correlation function for entangled operators is equal to correlator for the master--field algebra
$$
\lim_{\lambda\to 0}\langle a^{\varepsilon_1}_\lambda(t_1,k_1)\dots a^{\varepsilon_N}_\lambda(t_N,k_N)\rangle=\langle b^{\varepsilon_1}(t_1,k_1)\dots b^{\varepsilon_N}(t_N,k_N)\rangle,
$$
where the master field possesses relations with free statistics and is given by the sum of two contributions 
\begin{equation}\label{limit0}
b(t,k)=b_1(t,k)+b_2^{\dag}(t,k),
\end{equation}
independent in the free sense $b_1b_2^{\dag}=b_2b_1^{\dag}=0$ and relations 
\begin{equation}\label{limit1}
b_1(t,k)b_1^{\dag}(t',k')=2\pi\delta(t-t')\delta\left(\omega(k)+{\frac{1}{2}}\,k^2+kp\right)(N(k)+1)\delta(k-k'),
\end{equation}
\begin{equation}\label{limit2}
b_2(t,k)b_2^{\dag}(t',k')=2\pi\delta(t-t')\delta\left(\omega(k)-{\frac{1}{2}}\,k^2+kp\right)N(k)\delta(k-k').
\end{equation}
\begin{equation}\label{limit3}
b_1(t,k)p=(p+k)b_1(t,k),
\end{equation}
\begin{equation}\label{limit4}
b_2(t,k)p=(p-k)b_2(t,k).
\end{equation}

Correlation functions for the master--field are given by the Fock state for the algebra of free creations and annihilations with the generators $b_1$, $b_1^{\dag}$, $b_2$, $b_2^{\dag}$.
}

\medskip

We have to prove the deformed variation of the Wick theorem for the free statistics. Correlation function $\langle b^{\varepsilon_1}(t_1,k_1)\dots b^{\varepsilon_N}(t_N,k_N)\rangle$, $N=2n$ takes the form of a sum over all non-crossing diagrams (which follows from free statistics). Edges in the diagrams correspond to pairings $b_1b_1^{\dag}$, $b_2b_2^{\dag}$ (any edge $bb^{\dag}$ corresponds to pairing $b_1b_1^{\dag}$ and edge $b^{\dag}b$ corresponds to pairing $b_2b_2^{\dag}$). Formulae (\ref{limit1}), (\ref{limit2}) imply that any pairing of this form generates a term in the correlation function (\ref{N'-point}) (not taking in account the summation over $l$). The contribution inside the delta-function which contains summation over $l$ arises due to moving of the pairing through edges in the process of ordering of operator delta-functions taking in account relations (\ref{limit3}), (\ref{limit4}) (using the same arguments applied in the proof of Theorem 1 which generated similar contributions in correlation function (\ref{N'-point})).

\medskip

\noindent{\bf Acknowledgments.} This work is supported by the Russian Science Foundation under grant  19--11--00320.

\end{document}